\documentclass[11pt]{article}
\usepackage{hyperref}
\usepackage{amsmath,amsfonts,amssymb}
\usepackage{cite}
\usepackage{setspace}
\usepackage[top=2.45cm, bottom=2.5cm, left=2.45cm, right=2.45cm]{geometry}

\numberwithin{equation}{section}

\long\def\symfootnote[#1]#2{\begingroup%
\def\thefootnote{\fnsymbol{footnote}}\footnote[#1]{#2}\endgroup} 

\def\({\left (}
\def\){\right )}
\def\lb{\left [}
\def\rb{\right ]}
\def\lB{\left \{}
\def\rB{\right \}}

\def\Int#1#2{\int \textrm{d}^{#1} x \sqrt{|#2|}}
 
\def\Ket#1{\left|#1\right\rangle}
\def\BraKet#1#2{\left\langle#1|#2\right\rangle} 

\def\bbox{\bar{\Box}}

\def\wtil#1{\widetilde{#1}}
\def\ph#1{\phantom{#1}}

\def\ra{\rightarrow}

\def\p{\partial}

\def\sinh{\mathrm{sinh}}
\def\cosh{\mathrm{cosh}}
\def\tanh{\mathrm{tanh}}

\def\sech{\mathrm{sech}}
\def\csch{\mathrm{csch}}

\def\a{\alpha}
\def\b{\beta}
\def\g{\gamma}
\def\d{\delta}
\def\e{\epsilon}

\def\k{\kappa}
\def\l{\lambda}
\def\ll{\Lambda}
\def\n{\nabla}
\def\s{\sigma}
\def\t{\theta}

\def\z{\zeta}
\def\vp{\varphi}

\def\gg{\Gamma}

\def\H{{\cal H}}

\def\K{{\cal K}}
\def\L{{\cal L}}
\def\O{{\cal O}}

\begin{document}

\begin{titlepage}

\begin{center}

\ph{.}

\vskip 4 cm

{\Large \bf Nonlinear Dynamics of Parity-Even Tricritical Gravity\\in Three and Four Dimensions}

\vskip 1 cm

{Luis Apolo and Massimo Porrati}

\vskip .75 cm

{\em Center for Cosmology and Particle Physics, \\Department of Physics, New York University, \\4 Washington Place, New York, NY 10003, USA}

\end{center}

\vskip 1.25 cm

\begin{abstract}
\noindent Recently proposed ``multicritical'' higher-derivative gravities in Anti de Sitter space carry logarithmic representations of the Anti de Sitter isometry group. While generically non-unitary already at the quadratic, free-theory level, in special cases these theories admit a unitary subspace. The simplest example of such behavior is ``tricritical'' gravity. In this paper, we extend the study of parity-even tricritical gravity in $d=3, 4$ to the first nonlinear order. We show that the would-be unitary subspace suffers from a linearization instability and is absent in the full non-linear theory.
\end{abstract}

\end{titlepage}

\newpage


\section{Introduction}

Power-counting renormalizable theories of gravity can be obtained by adding to the Einstein-Hilbert action appropriate terms, quadratic in the Ricci and Weyl tensors. In the absence of a cosmological constant, these theories admit Minkowski space as a background, but they are also perturbatively non-unitary~\cite{Stelle:1976gc, Stelle:1977ry}. Since the whole point of having a power-counting renormalizable theory of gravity is to make perturbative calculations possible, these theories have been justly abandoned for many years\footnote{Theories with an infinite number of derivatives are inherently nonlocal. As shown in ref.~\cite{ArkaniHamed:2002fu} in this case one faces a classic ``rob Peter to pay Paul'' dilemma: either introducing ghosts or giving up causality.}. A recent resurgence in interest towards these theories was triggered by the study of quadratic-curvature actions {\em with} cosmological constant in four~\cite{Lu:2011zk} and $d$~\cite{Deser:2011xc} dimensions. In either case, it was found that there exists a choice of parameters for which these theories possess one AdS background on which neither massive fields, nor massless scalars or vectors propagate. Moreover, on the AdS background the standard graviton, i.e. the massless tensor mode of Einstein-Hilbert gravity, also propagates and has vanishing energy~\cite{Lu:2011zk,Deser:2011xc}.

Besides those that satisfy the homogeneous Einstein equations on AdS$_d$, other tensor modes propagate in the ``critical'' theory~\cite{Alishahiha:2011yb,Gullu:2011sj,Bergshoeff:2011ri}. Their asymptotic behavior at space-like infinity differs from standard Einstein-Hilbert modes by terms logarithmic in the AdS radial coordinate. A complete set of propagating modes for four-dimensional critical gravity was presented in~\cite{Bergshoeff:2011ri}.

Critical gravities are also interesting in the light of the AdS/CFT duality~\cite{Maldacena:1997re,Gubser:1998bc,Witten:1998qj}. Indeed, they were conjectured to be dual to logarithmic conformal field theories (LCFTs)~\cite{Grumiller:2008qz,Skenderis:2009nt,Grumiller:2009mw,Grumiller:2009sn, Alishahiha:2010bw,Liu:2009pha, Johansson:2012fs}. Although typically non-unitary, LCFTs have found applications in condensed matter physics, where they are used in the study of e.g.~critical phenomena, turbulence and percolation. As such, critical gravities might represent gravitational duals of certain strongly coupled condensed matter systems.

One could also try to make sense of critical gravities as toy models for quantum gravity. Then, however, one has to deal with the non-unitarity of these theories. Note that the logarithmic modes, which are responsible for the violation of unitarity, obey {\em different} boundary conditions than the original massive gravitons. It has been proposed that by imposing strict Brown-Henneaux boundary conditions one could get rid of the problematic logarithmic modes and obtain a theory that is possibly unitary. This approach has been taken recently in two particular three-dimensional higher-derivative gravity models in AdS: Topologically Massive Gravity (TMG) \cite{Deser:1981wh,Deser:1982vy} and New Massive Gravity (NMG) \cite{Bergshoeff:2009hq,Bergshoeff:2009aq}. Imposing Brown-Henneaux boundary conditions on critical TMG leads to so-called chiral gravity, that is dual to a two-dimensional chiral CFT \cite{Li:2008dq,Maloney:2009ck}. In spite of an apparent non-unitarity at the linear level~\cite{Grumiller:2008qz,Giribet:2008bw} the theory admits a chiral, unitary subsector at the classical level~\cite{Maloney:2009ck}. TMG however cannot be generalized to higher dimensions; the chiral splitting into right and left movers is unique to a two-dimensional boundary.

New Massive Gravity instead can also be formulated in dimensions higher than three. At the critical point it becomes a higher-dimensional critical gravity \cite{Lu:2011zk,Deser:2011xc,Alishahiha:2011yb,Bergshoeff:2011ri,Porrati:2011ku}.  In that theory, by imposing strict Brown-Henneaux boundary conditions in $d>3$, one obtains a theory which describes a massless graviton with zero on-shell energy and black holes with zero mass and entropy \cite{Lu:2011zk,Liu:2009bk, Liu:2009kc,Lu:2011ks}. Modding out these states leaves the vacuum as the only physical state \cite{Porrati:2011ku}.

The aim of imposing specific boundary conditions is to obtain a consistent unitary truncation of the full non-unitary critical theory. On the dual field theory side this means that there should exist a consistent truncation of the LCFT that leads to an ordinary CFT. LCFTs are characterized by the fact that there are fields with degenerate scaling dimensions on which the Hamiltonian acts non-diagonally~\cite{Gurarie:1993xq,Flohr:2001tj,Flohr:2001zs}. These degenerate fields form so-called Jordan cells. One of the fields in a Jordan cell corresponds to a zero norm state, while the other fields are referred to as logarithmic partners. The rank of the LCFT then refers to the dimensionality of the Jordan cell. The LCFTs dual to critical gravities have rank 2 and operators thus have one logarithmic partner. The truncation mentioned above then corresponds to truncating these logarithmic partners.

In~\cite{Bergshoeff:2012sc}, holographic scalar LCFTs of rank $r > 2$ were studied, in which the bulk side was made of a scalar field in a fixed AdS background with higher derivatives up to order $2r$. Ref.~\cite{Bergshoeff:2012sc} shows that at the quadratic level, i.e. without introducing interactions, the dual field theory of an {\em odd-rank} LCFT allows for a truncation to a unitary CFT. This truncation is different from the one in the rank 2 case mentioned above in the sense that it keeps modes that correspond to the null state plus half of the logarithmic modes, whereas in the usual (rank 2) critical gravity proposals the single logarithmic mode is truncated. Ref.~\cite{Bergshoeff:2012ev} extended the study of $r=3$ theories to three-dimensional parity-even gravity. It showed that in those theories there exists a charge that vanishes for all but the $\log^2(z)$ modes. So, the truncation to a physical, positive-norm subspace could possibly be enforced by setting that conserved charge to zero. 

In this paper we re-examine the question of conserved charges and unitary truncations in the specific case of parity-even tricritical gravity in three and four dimensions. Its linearized action was given e.g. in~\cite{Nutma:2012ss}.

When extending the analysis of tricritical gravity to nonlinear order, we are faced with the non-uniqueness of the action: many different actions reduce to the same at quadratic order in the fluctuation ($h_{\mu\nu}$) of the metric ($g_{\mu\nu}$) around the AdS$_d$ background ($\bar{g}_{\mu\nu}$), while differing at order $h_{\mu\nu}^3$. Yet a few general structures emerge.

First of all, let us write the equations of motion to quadratic order in terms of a set of fluctuations  around the AdS background, denoted by $(h_{\mu\nu},f_{\mu\nu},k_{\mu\nu})$. The first is the fluctuation of the metric, the last is nonzero only if the metric has an asymptotic behavior $h_{\mu\nu} \sim z^2\log^2(z)$ near the boundary, while $f_{\mu\nu}$ is the would be physical mode, i.e. the mode that spans a positive Hilbert space when $k_{\mu\nu}=0$. In a schematic form the equations of motion are:
  \begin{align}
  L_{\mu\nu}^{\phantom{\mu\nu}\rho\s}h^{(2)}_{\rho\s} & = f_{\mu\nu}^{(2)} + Q_1[h^{(1)},f^{(1)},k^{(1)}], \notag \\
  L_{\mu\nu}^{\phantom{\mu\nu}\rho\s}f^{(2)}_{\rho\s} & = k_{\mu\nu}^{(2)} + Q_2[h^{(1)},f^{(1)},k^{(1)}], \notag \\
  L_{\mu\nu}^{\phantom{\mu\nu}\rho\s}k^{(2)}_{\rho\s} & = Q_3[h^{(1)},f^{(1)},k^{(1)}]. \label{eq:charges}
  \end{align}
These equations mean that the fluctuations to second order are determined by the linearized Einstein equation ($L_{\mu\nu}^{\phantom{\mu\nu}\rho\s}h_{\rho\s}$) sourced by linear and quadratic terms. The fields inside the quadratic terms $Q_1,Q_2,Q_3$ are solutions to the linear equations of motion. A first thing to notice is that only $k^{(2)}_{\mu\nu}$ is sourced purely by quadratic terms. This is important because then the General Relativity version of Gauss's law can be applied to extract a conserved quantity that is at the same time: a) a boundary integral of a linear function of $k^{(2)}_{\mu\nu}$ and b) a bulk integral that is a quadratic function of {\em solutions to the linearized equations of motion}.

So, one can find a Gauss law charge which, for each Killing vector of the background metric, $\xi$, associates a unique expression given by a surface integral. The  surface integral form of the charge is given by
  \begin{align}
  Q[\xi] = \oint \textrm{d}S_i \sqrt{|\bar{g}|} \( \bar{\n}_{\s} \K^{0 i \nu\s} \xi_\nu- \K^{0 j \nu i} \bar{\n}_j \xi_\nu \),
  \end{align}
where $\K_{\mu\rho\nu\s}$ depends linearly on $k^{(2)}_{\mu\nu}$. This equation makes evident that the charge vanishes unless $k^{(2)}_{\mu\nu} \neq 0$, i.e. unless -- as shown in Section 3 -- the metric has a $z^2\log^2(z)$ behavior near the boundary. On the other hand eqs.~\eqref{eq:charges} show that the conserved charge $Q[\xi]$ can also be given a volume integral representation. One obtains an expression of the form
  \begin{align}
  Q[\xi]= \Int{d-1}{\bar{g}} \bar{g}^{00}Q_{3\, 0\mu}\xi^\mu.
  \end{align}

The crucial point made in this paper is that the expression for $Q_{3\,\mu\nu}$ we find has the following two properties:
  \begin{enumerate}
  \item $Q_{3\, \mu\nu}$ is generically not positive definite; therefore, a fluctuation with nonzero $k_{\mu\nu}$ and $h_{\mu\nu}$ in the bulk can, nevertheless, produce a metric fluctuation behaving as $h\sim z^2(\log(z) + \textrm{constant})$ near the boundary. Thus, while it is consistent to define a physical subspace by setting to zero the charge $Q[\xi]$, that subspace can still contain states of negative norm.
  \item More dramatically, on the would-be physical subspace, where $k_{\mu\nu} = 0$ throughout the bulk, $Q[\xi]$ is {\em negative definite} if $f^{(1)}_{\mu\nu} \neq 0$. But then, by the Gauss law implied by eqs.~\eqref{eq:charges}, $k^{(2)}_{\mu\nu} \neq 0$ unless $f^{(1)}_{\mu\nu} = 0$.
  \end{enumerate}
So, instead of selecting a physical subspace, the conserved charge $Q[\xi]$ says that the only subspace with $k_{\mu\nu} = 0$ up to quadratic order, contains no fluctuations of type $f^{(1)}_{\mu\nu}$. But they are precisely the fluctuations that span the physical Hilbert space! Moreover, we point out that in three dimensions the norm of the $f_{\mu\nu}$ modes is identical to the Einstein Gravity norm. Since only vector $f_{\mu\nu}=D_{(\mu}A_{\nu)}$ modes exist there, their scalar  product vanishes anyway, as it does in Einstein Gravity.  So we arrive at the main result of our paper:
  \begin{center}
  {\bf Parity-even tricritical gravity in \emph{d} = 3, 4 has no physical, positive-metric subspace.}
  \end{center}
The apparent subspace found in the free theory is an artifact of linearizing the equations of motion, a possibility already hinted at in ref.~\cite{Bergshoeff:2012ev}.

The rest of the paper is devoted to derive and make precise the formulas given above. In particular, Section 2 reviews the construction while Section 3 presents the modes of tricritical gravity. Section 4 shows that at linear order in the equations of motion it is possible to truncate the theory to a unitary subsector. Section 5 constructs the bulk and boundary expressions for the conserved charges in the simplest realization of tricritical gravity. Section 5 also shows explicitly the negative definiteness of the conserved charge on the subspace $k^{(1)}_{\mu\nu} = 0$, thus proving the main result of the paper for that particular realization of tricritical gravity. Section 6 extends the result to generic tricritical gravities, that is theories that may differ from the simplest realization of such theory by terms cubic or higher order in the fluctuations $(h,f,k)$. We end with our conclusions in Section 7, where we also notice that the non-positivity of the $Q[\xi]=0$ Hilbert space is also a problem for multicritical gravities.


\section{The action of tricritical gravity}

\subsection{Minimal action}
In this section we construct the simplest realization of parity-even tricritical gravity in $d > 2$ dimensions. To establish some conventions and notation we begin by considering the Einstein-Hilbert action with a cosmological constant $\ll$,
  \begin{align}
  S = \frac{1}{8\pi\k} \Int{d}{g} \( R - 2\ll \).  \label{eq:einsteinhilbert}
  \end{align}
The equations of motion are given by
  \begin{align}
  H_{\mu\nu} = R_{\mu\nu} - \frac{1}{2}Rg_{\mu\nu} +\ll g_{\mu\nu} = 0,
  \end{align}
and admit an AdS solution with radius of curvature $\ell$ where $\ll = - {(d-2)(d-1) \over 2 }\,\ell^{-2}$. In the AdS background $\bar{g}_{\mu\nu}$ the curvature tensors read
  \begin{align}
  \bar{R} = d\l, && \bar{R}_{\mu\nu} = \l \bar{g}_{\mu\nu}, && \bar{R}^{\s}_{\ph{\s}\mu\rho\nu} = \frac{\l}{d-1} \(\d^\s_\rho \,\bar{g}_{\mu\nu} - \d^\s_\nu \,\bar{g}_{\mu\rho} \),  \label{eq:curvatures}
  \end{align}
where for convenience we have defined the {\it reduced} cosmological constant
  \begin{align}
  \l = \frac{2}{d-2}\,\ll.
  \end{align}
By definition the tensor $H_{\mu\nu}$ vanishes in an AdS background and is covariantly conserved thanks to the Bianchi identity. These properties make $H_{\mu\nu}$, rather than $R_{\mu\nu}$ or $R$, useful in the construction of the action. 

The simplest action for tricritical gravity quadratic in $H_{\mu\nu}$ may be written as 
  \begin{align}
  S_3 = -\frac{\k}{2} \Int{d}{g} \lb \a_1 \(\n_\s H_{\mu\nu}\)^2 + \b_1 \( \n_\s H \)^2 - \a_o\l H_{\mu\nu}^2 - \b_o\l H^2 \rb,  \label{eq:tricriticalaction}
  \end{align}
where $H = g^{\mu\nu}H_{\mu\nu}$, $\k$ is a positive constant with dimensions of mass$\,^{d-6}$, and we set $\a_1 = 1$. As mentioned above this action automatically admits an AdS solution given by eqs.~\eqref{eq:curvatures}. Note that the action does not contain terms squared in the Riemann tensor. In three and four dimensions this term may be omitted since the Gauss-Bonnet term
  \begin{align}
  S_{\textrm{GB}} = \Int{d}{g} \( R_{\mu\s\nu\rho}^2 - 4R_{\mu\nu}^2 + R ^2 \),  \label{eq:gaussbonnet}
  \end{align}
does not contribute to the classical equations of motion. For $d > 4$ the addition of this term leads to a more general class of critical gravities, examples of which have been considered in \cite{Deser:2011xc,Nutma:2012ss}. Notice that we have also omitted the following quadratic terms
  \begin{align}
  \(\n_\a R_{\mu\s\nu\rho}\)^2, && \n^\mu\n^\nu R_{\mu\s\nu\rho}R^{\s\rho}, && \n^{\mu} R_{\mu\nu} \n^{\nu} R,
  \end{align}
since, as pointed out in \cite{Myers:2010ru}, these can be written in terms of $(\n_\s H_{\mu\nu})^2$, $(\n_\s H)^2$, and total derivatives using the Bianchi identities. In eq.~\eqref{eq:tricriticalaction} we have also not included the Lorentz-Chern-Simons term \cite{Deser:1981wh,Deser:1982vy} which in three dimensions yields a tricritical gravity of odd parity \cite{Bergshoeff:2012ev}. Finally let us note that it is possible to add to the action another term proportional to \eqref{eq:einsteinhilbert} since the basis $H_{\mu\nu}^2$, $H^2$ is not linearly independent. 
  %
  %
This term leads away from criticality, i.e. towards points where the gravitons are no longer {\it massless} in the AdS sense, and with the help of hindsight we omit it from the start.

The equations of motion that follow from the action \eqref{eq:tricriticalaction} are given by
  \begin{align}
  0 & = T_{\mu\nu} + \frac{1}{\sqrt{|g|}} \frac{\d H_{\rho\s}}{\d g_{\mu\nu}}\frac{\d S}{\d H_{\rho\s}}, \hskip 4cm T_{\mu\nu} = -\frac{1}{\k} \frac{1}{\sqrt{|g|}} \frac{\d S}{\d g^{\mu\nu}}, \label{eq:tmunu} \\
  0 & = T_{\mu\nu} + \frac{1}{2} \( \wtil{R}_{\mu\nu}g^{\rho\s} - \wtil{R} \d^\rho_\mu\,\d^\s_\nu +  2L_{\mu\nu}^{\ph{\mu\nu}\rho\s} \) \Big [ \big ( \Box + \a_o \l \big ) H_{\rho\s} + \big ( \b_1 \Box + \b_o \l \big )g_{\rho\s}H \Big]. \label{eq:nonlineareom}
  \end{align}
where the {\em effective stress-energy tensor} $T_{\mu\nu}$ does not contain variations of $H_{\mu\nu}$ with respect to the metric. Here the operator $L_{\mu\nu}^{\ph{\mu\nu}\rho\s}$ is equal to the linearized Einstein tensor given by
  \begin{align}
  \begin{split}
  L_{\mu\nu}^{\ph{\mu\nu}\rho\s} = \frac{1}{2} & \( \n^\rho \n_\mu \,\d_\nu^\s + \n^\rho \n_\nu \,\d_\mu^\s - \Box \,\d_\mu^\rho \,\d_\nu^\s - \n_\mu \n_\nu g^{\rho\s} - g_{\mu\nu}\n^\rho \n^\s + g_{\mu\nu}\, \Box \,g^{\rho\s} \right. \\
  & \left. + \l\, g_{\mu\nu}\,g^{\rho\s} - 2 \l \,\d_\mu^\rho \,\d_\nu^\s \),  \label{eq:Hlinear}
  \end{split}
  \end{align}
and any curvature tensor with a tilde vanishes in the background, e.g. $\wtil{R} = R - \bar{R}$. Written in this way it is clear that only the term proportional to $L_{\mu\nu}^{\ph{\mu\nu}\rho\s}$ contributes to the linearized equations of motion. Letting $g_{\mu\nu} \ra \bar{g}_{\mu\nu} + h_{\mu\nu}$ these read
  \begin{align}
  0 = \( \bar{L}_{\mu\nu}^{\ph{\mu\nu}\a\b} \,\bbox\, \bar{L}_{\a\b}^{\ph{\a\b}\rho\s} + \b_1 \bar{L}_{\mu\nu\a}^{\ph{\mu\nu}\a}  \,\bbox\, \bar{L}_{\b}^{\b\rho\s} + \a_o \l \bar{L}_{\mu\nu}^{\ph{\mu\nu}\a\b} \bar{L}_{\a\b}^{\ph{\a\b}\rho\s}  + \b_o \l \bar{L}_{\mu\nu\a}^{\ph{\mu\nu}\a} \bar{L}_{\b}^{\b\rho\s} \) h_{\rho\s}.  \label{eq:linearizedeom}
  \end{align}
Next we fix the coefficients in \eqref{eq:tricriticalaction} so that no scalar modes propagate and the masses of all gravitons are degenerate and equal to their AdS value, $m^2 = -2\ell^{-2}$.  In the gauge $\bar{\n}^{\mu}h_{\mu\nu} = \bar{\n}_\nu\, h$ the trace of the linearized equations of motion yields
  \begin{align}
  0 = & \lb 1 + (d-1)\b_1 \rb \l\, \bbox^2\, h + \lb \frac{d+1}{d-1} + d\b_1 + \a_o + (d-1) \b_o \rb \l^2 \,\bbox\, h + \( \a_o + d \b_o \) \l^3 h.
  \end{align}
It is therefore possible to remove the scalar mode by requiring that 
  \begin{align}
  \b_1 = -\frac{1}{d-1}, && \b_o = -\frac{1}{d-1}\( \a_o + \frac{1}{d-1} \), && \a_o + d \b_o \ne 0.  \label{eq:tricriticalcoeff}
  \end{align}
With the perturbation $h_{\mu\nu}$ now transverse and traceless, the second and fourth terms in eq.~\eqref{eq:linearizedeom} drop out and the equations of motion become
  \begin{align}
  0 = & \bbox^3\,h_{\mu\nu} + \( \a_o - \frac{4}{d-1} \) \l\,\bbox^2\,h_{\mu\nu} + \frac{4\lb 1-\a_o (d-1) \rb}{(d-1)^2}\,\l^2\,\bbox\,h_{\mu\nu} + \frac{4\, \a_o}{(d-1)^2}\, \l^3 h_{\mu\nu}.
  \end{align}
Besides the massive spin-1 modes to be discussed in the next section, this equation describes three degenerate spin-2 fields with mass $m^2 = -2\ell^{-2}$ if
  \begin{align}
  \a_o = -\frac{2}{d-1}. \label{eq:tricriticalcoeff2}
  \end{align}
Hence at the critical point given by eqs.~\eqref{eq:tricriticalcoeff} and \eqref{eq:tricriticalcoeff2} the scalar mode disappears from the perturbative spectrum and the linearized equations of motion are given by 
  \begin{align}
  0 = \( \bbox + 2\ell^{-2} \)^3 h_{\mu\nu}.  \label{eq:eom}
  \end{align}

The action of the simplest, parity-even tricritical gravity in $d$ dimensions is thus
  \begin{align}
  S_3 = -\frac{\k}{2} \Int{d}{g} \lb \(\n_\s H_{\mu\nu}\)^2 - \frac{1}{d-1} \( \n_\s H \)^2 + \frac{2}{d-1} \l H_{\mu\nu}^2 - \frac{1}{(d-1)^2} \l H^2 \rb.  \label{eq:tricriticalaction2}
  \end{align}
  %
  %
  %
The action may be expressed more conveniently in terms of $R_{\mu\nu}$ and $R$, where we see explicitly that the Einstein-Hilbert term has a positive sign
  \begin{align}
  \begin{split}
  S_3 = -\frac{\k}{2} \Int{d}{g} & \lb \(\n_\s R_{\mu\nu}\)^2 - \frac{d}{4(d-1)} \( \n_\s R \)^2 + \frac{2}{d-1}\, \l R_{\mu\nu}^2 + \frac{d^2-6d+4}{4(d-1)^2}\, \l R^2 \right. \\
  & \left. - \frac{(d-2)^3}{2(d-1)^2} \,\l^2 \( R- \frac{d}{2}\,\l \)\rb.  
  \end{split}
  \end{align}
Written in this form it is possible to check that for $d = 3$ one obtains the parity-even tricritical gravity studied in \cite{Bergshoeff:2012ev}. In four dimensions any action for tricritical gravity must either contain these terms -- plus eventually other terms that do not contribute to the linearized equations of motion -- or  be reducible into such a form. As an illustration of this consider the four-dimensional, Weyl-invariant tricritical action constructed in \cite{Nutma:2012ss} and written there as
  \begin{align}
  S_W = \frac{\k}{2} \Int{4}{g} \lb \frac{1}{2} \, C_{\mu\rho\nu\s}\,\Box\,C^{\mu\rho\nu\s} - \frac{2}{3}\l\, C_{\mu\rho\nu\s}^2 + \frac{4}{9} \l^2 (R-2\l) \rb,
  \end{align}
where $C_{\mu\rho\nu\s}$ is the Weyl tensor. This action is up to total derivatives equivalent to 
  \begin{align}
  S_W = S_3 + \frac{\k}{6}\,\l S_{GB} +  \O(\wtil{R}^{3}),
  \end{align}
where $S_{GB}$ is the Gauss-Bonnet term and $\O(\wtil{R}^3)$ consists of terms cubic in curvature tensors that vanish in the background and do not contribute to the linearized equations of motion.

\subsection{Auxiliary fields}

The action of tricritical gravity can be written in a form that contains at most two derivatives in terms of appropriate auxiliary fields. This will be useful in finding the solutions to the linearized equations of motion, and in the definition of the inner product given in the next section. The action with auxiliary fields can be obtained from inspection of the equations of motion \eqref{eq:nonlineareom} and is given by\footnote{In the remainder of the paper the coefficients $(\a_n,\b_n)$ are taken at the critical point given by eqs.~\eqref{eq:tricriticalcoeff} and \eqref{eq:tricriticalcoeff2} but we avoid explicit expressions to reduce clutter.}
  \begin{align}
  \begin{split}
  S_3 = -\frac{\k}{2} \int d^dx \sqrt{|g|} & \lB \(\n_\s F_{\mu\nu}\)^2 + \g_1 \( \n_\s F \)^2 - \a_o\l F_{\mu\nu}^2 - \g_o\l F^2 \right. \\ 
  & + 4 \lb K^{\mu\nu}H_{\mu\nu} - \( K^{\mu\nu}F_{\mu\nu} - KF \) \rb \Big \},  \label{eq:auxiliaryaction}
  \end{split}
  \end{align}
where $F = g^{\mu\nu}F_{\mu\nu}$, $K = g^{\mu\nu}K_{\mu\nu}$, and $F_{\mu\nu}$ has dimensions of mass$\,^{2}$ while $K_{\mu\nu}$ has dimensions of mass$\,^{4}$. The constants $\g_n$ are given in terms of $\a_n$ and $\b_n$ by
  \begin{align}
  \g_n = (d-2)\,\a_n + (d-1)^2 \b_n,
  \end{align}
and we recall that $\a_1 = 1$. The equations of motion that follow from variation with respect to $K_{\mu\nu}$, $F_{\mu\nu}$, and $g_{\mu\nu}$ are, respectively,
  \begin{align}
  H_{\mu\nu} & = F_{\mu\nu} - g_{\mu\nu}F,  \label{eq:aux1} \\
  -\frac{1}{2} \lb \( \Box + \a_o \l\) F_{\mu\nu} + \( \g_1 \Box + \g_o \l \) g_{\mu\nu} F \rb & = K_{\mu\nu} - g_{\mu\nu}K,  \label{eq:aux2} \\
  \( \wtil{R}_{\mu\nu}g^{\rho\s} - \wtil{R} \d^\rho_\mu\,\d^\s_\nu +  2 L_{\mu\nu}^{\ph{\mu\nu}\rho\s} \) K_{\rho\s} & =  T_{\mu\nu},  \label{eq:aux3}
  \end{align}
where $T_{\mu\nu}$ is defined in \eqref{eq:tmunu} and does not contain the variation of $H_{\mu\nu}$ with respect to the metric. It is easy to show that these equations yield back the equations of motion \eqref{eq:nonlineareom}, while substituting them in \eqref{eq:auxiliaryaction} restores the action to its original form \eqref{eq:tricriticalaction}.

Let us now consider the linearized equations of motion.  Since $H_{\mu\nu}$ vanishes in the background, eqs.~\eqref{eq:aux1} and \eqref{eq:aux2} imply that $\bar{F}_{\mu\nu}$ and $\bar{K}_{\mu\nu}$ must also vanish. To linear order the fields may be expanded as follows
  \begin{align}
  g_{\mu\nu} \ra \bar{g}_{\mu\nu} + h_{\mu\nu}, && F_{\mu\nu} \ra f_{\mu\nu}, && K_{\mu\nu} \ra k_{\mu\nu}, 
  \end{align}
and the linearized equations of motion are
  \begin{align}
  \bar{L}_{\mu\nu}^{\ph{\mu\nu}\rho\s} h_{\rho\s} & = f_{\mu\nu} - \bar{g}_{\mu\nu} f, \label{eq:laux1} \\  
  -\frac{1}{2} \lb \( \bbox + \a_o \l\) f_{\mu\nu} + \( \g_1 \bbox + \g_o \l \) \bar{g}_{\mu\nu} f \rb & = k_{\mu\nu} - \bar{g}_{\mu\nu}k,  \label{eq:laux2} \\
  \bar{L}_{\mu\nu}^{\ph{\mu\nu}\rho\s}k_{\rho\s} & = 0.   \label{eq:laux3}
  \end{align}
The first equation tells us that $f_{\mu\nu}$ is gauge-invariant under linearized diffeomorphisms $\d h_{\mu\nu} = \bar{\n}_{(\mu}\xi_{\nu)}$, since $\bar{L}_{\mu\nu}^{\ph{\mu\nu}\rho\s}h_{\rho\s}$ is gauge-invariant, and from the second we learn that $k_{\mu\nu}$ must be gauge-invariant too. In the gauge $\bar{\n}^\mu h_{\mu\nu} = \bar{\n}_\nu\, h$ the vanishing trace of eq.~\eqref{eq:linearizedeom} and eqs.~\eqref{eq:laux1}, \eqref{eq:laux2} imply the fields are traceless,
  \begin{align}
  0 = h = k = f,  \label{eq:traceless}
  \end{align}
where $\vp = \bar{g}^{\mu\nu} \vp_{\mu\nu}$ for any $\vp_{\mu\nu}$; while the Bianchi identity and eqs.~\eqref{eq:laux1}, \eqref{eq:laux2} imply the auxiliary fields are transverse,
  \begin{align}
  0 =\bar{\n}^\mu h_{\mu\nu} = \bar{\n}^\mu f_{\mu\nu} = \bar{\n}^\mu k_{\mu\nu}.  \label{eq:transverse}
  \end{align}
Using $\a_o = -2/(d-1)$ the gauge-fixed, linearized equations of motion of the auxiliary fields are, not surprisingly,
  \begin{align}
  -\frac{1}{2} \( \bbox + 2\ell^{-2}\)h_{\mu\nu} & = f_{\mu\nu} \label{eq:llaux1} \\
  -\frac{1}{2} \( \bbox + 2\ell^{-2}\)f_{\mu\nu} & = k_{\mu\nu}  \label{eq:llaux2} \\
  -\frac{1}{2} \( \bbox + 2\ell^{-2}\)k_{\mu\nu} & = 0.  \label{eq:llaux3}
  \end{align}
  %


\section{Modes of tricritical gravity}
In this section we present the solutions to the linearized equations of motion of tricritical gravity in four dimensions. A similar analysis in three dimensions has been recently performed in \cite{Bergshoeff:2012ev}. 

In \cite{Bergshoeff:2011ri} the solutions to (bi)critical gravity, given by the last two equations above, were found in terms of highest weight representations of the isometry group $SO(2,3)$ of AdS$_4$. From these all possible solutions can be obtained by acting with the negative root generators of the algebra. To linear order, the auxiliary fields of critical gravity, like those of tricritical gravity, are gauge-invariant under linearized diffeomorphisms, transverse, and traceless. This means that the results of ref.~\cite{Bergshoeff:2011ri}, which we now summarize, can be easily generalized to critical gravities of higher rank. 

In global coordinates the metric of AdS$_4$ may be written as
  \begin{align}
  ds^2 = \ell^2 \lb -\cosh^2(r)\, dt^2 + dr^2 + \sinh^2(r)\( d\t^2 + \sin^2(\t)\,d\phi^2 \) \rb,  \label{eq:globalmetric}
  \end{align}
where $r$ is the radial coordinate. The isometry group of AdS$_4$ can be decomposed into two Cartan generators $H_1$, $H_2$, four positive and four negative root generators $E^{\pm\a_n}$, $n = 1\dots4$. The Killing vectors corresponding to the Cartan generators are given by
  \begin{align}
  H_1 = i\p_t, && H_2 = -i\p_\phi,
  \end{align}
while the Killing vectors corresponding to the positive and negative root generators $E^{\pm\a_2}$ are
  \begin{align}
  E^{\a_2} = -i e^{i\phi}\p_\t + e^{i\phi}\cot(\t)\p_\phi, && E^{-\a_2} = -(E^{\a_2})^*.
  \end{align}
Clearly $H_1$ is the generator of time translations and it commutes with $H_2$ and $E^{\pm\a_2}$, which are the generators of rotations, i.e.
  \begin{align}
  & [H_1,H_2] = 0, && [H_1,E^{\pm\a_2}] = 0,  \label{eq:algebra1} \\
  & [H_2,E^{\pm\a_2}] = \pm E^{\pm\a_2},  && [E^{\a_2},E^{-\a_2}] = 2 H_2.  \label{eq:algebra2}
  \end{align}
These operators generate the maximal compact subgroup of $SO(2,3)$, namely $SO(2)\times SO(3)$. When the generators of $SO(2,3)$ act on a highest weight state $\psi_{\mu\nu}$ we have
  \begin{align}
  \L_{H_1} \psi_{\mu\nu} = E_o \psi_{\mu\nu}, && \L_{H_2} \psi_{\mu\nu} = s\, \psi_{\mu\nu}, && \L_{E^{\a_n}} \psi_{\mu\nu} = 0, \quad n = 1,\dots,4,
  \end{align}
where $\L_\xi$ is the Lie derivative along $\xi$, $s$ is the helicity, and $E_o(E_o - 3) = m^{2}\ell^{2} + 2$ where $m^{2}$ is the mass of the graviton. For a \emph{massless} graviton in AdS, i.e. one propagating only two degrees of freedom, $m^{2} = -2\ell^{-2}$ and therefore $E_o = 3$.

With these basic ingredients in place let us now consider the solutions to the linearized equations of motion starting with eq.~\eqref{eq:llaux3}. With $k_{\mu\nu}$ transverse and traceless, a highest weight solution for $s = 2$ was found in \cite{Bergshoeff:2011ri} which reads, 
  \begin{align}
  \psi_{tt} & = - \psi_{t\phi} = \psi_{\phi\phi} = \vp  && \psi_{\t\t} = -\cot^2(\t)\,\vp \notag \\
  \psi_{tr} & = - \psi_{r\phi} = i\,\csch(r)\,\sech(r)\,\vp  && \psi_{t\t} = - \psi_{\t\phi} = i\,\cot(\t)\,\vp \label{eq:einsteinmodes1} \\
  \psi_{rr} & = -4\,\csch^2(2r)\,\vp  && \psi_{r\t} = -2\,\cot(\t)\,\csch(2r)\vp \notag  
  \end{align}
where
  \begin{align}
  \vp = e^{-iE_o t + 2i\phi}\,\sin^2(\t)\, \sinh^{1-E_o/2}(2r)\,\tanh^{1+E_o/2}(r).  \label{eq:einsteinmodes2}
  \end{align}
From eqs.~\eqref{eq:algebra1} and \eqref{eq:algebra2} we see that $E^{-\a_2}$ lowers the helicity of this solution without changing its energy $E_o$. Thus acting with $\L_{E^{-\a_2}}$ repeatedly on $\psi_{\mu\nu}$ yields the five possible helicities of a spin-2 mode. With $E_o = 3$ the helicity $\pm 2$ solutions correspond to a massless graviton in AdS, i.e. to an Einstein mode that we denote by $\psi^{E}_{\mu\nu}$; whereas the spin-1 solutions correspond to modes of the form $\bar{\n}_{(\mu}A_{\nu)}$ which cannot be gauged away since $k_{\mu\nu}$ is gauge-invariant under linearized diffeomorphisms. Consistency with eqs.~\eqref{eq:traceless} and \eqref{eq:transverse} dictates that $A_\mu$ must satisfy
  \begin{align}
  \( \bbox - 3\ell^{-2} \) A_\mu = 0, 
  \end{align}
which in AdS$_4$ corresponds to a massive\footnote{Recall that a {\it massless} vector with only two degrees of freedom has $m^2 = -3\ell^{-2}$ instead.} spin-1 field in the $D(4,1)$ representation of $SO(2,3)$. Acting with the other negative weight generators of $SO(2,3)$ on $\psi_{\mu\nu}$ yields the descendant states and all possible solutions to the equations of motion.

Let us now look at the highest weight solutions to eq.~\eqref{eq:llaux2} which may be more conveniently written as
  \begin{align}
  \( \bbox + 2\ell^{-2} \)^2 f_{\mu\nu} = 0.
  \end{align}
The solution to eq.~\eqref{eq:llaux3} given in the previous paragraph is also a solution to this equation. As shown in \cite{Grumiller:2008qz, Ghezelbash:1998rj, Kogan:1999bn}, another class of solutions, logarithmic in the coordinates, are given by 
  \begin{align}
  \psi^{\log}_{\mu\nu} = \frac{c\ell^2}{2E_o - 3}\,\frac{\p \psi_{\mu\nu}}{\p E_o}\bigg |_{E_o = 3},
  \end{align}
where $c$ is a constant. With $\psi_{\mu\nu}$ given by eqs.~\eqref{eq:einsteinmodes1} and \eqref{eq:einsteinmodes2} we obtain
  \begin{align}
  f_{\mu\nu} = \frac{2\ell^2}{3} \lB it + \log\lb \sqrt{2}\,\cosh(r)\rb \rB \psi_{\mu\nu},
  \end{align}
where we have chosen $c = -2$ so that eq.~\eqref{eq:llaux2} holds. 

This solution is transverse and traceless as required by the auxiliary field $f_{\mu\nu}$. Near the boundary of AdS it behaves as $f_{\mu\nu} \sim r e^{-r}$, or alternatively as $f_{\mu\nu} \sim z^2\log(z)$ in Poincar\'{e} coordinates. By acting repeatedly with the lowering operator $E^{-\a_2}$ we obtain a tower of states with all possible helicities. We thus have log states with helicity $\pm 2$ that arise from eq.~\eqref{eq:llaux2} when $k_{\mu\nu} = \psi^{E}_{\mu\nu}$, and log states with spin 1 obtained when $k_{\mu\nu} = \bar{\n}_{(\mu}A_{\nu)}$.

Finally let us consider the highest weight solutions to eq.~\eqref{eq:llaux1} or equivalently, to the full equations of motion given by
  \begin{align}
  \( \bbox + 2\ell^{-2} \)^3 h_{\mu\nu} = 0.
  \end{align}
Clearly $\psi_{\mu\nu}$ and $\psi^{\log}_{\mu\nu}$ are also solutions to this equation. As is to be expected there are other solutions logarithmic in the coordinates given by
  \begin{align}
    \psi^{\log^2}_{\mu\nu} = \frac{c'\ell^2}{2E_o - 3}\,\frac{\p}{\p E_o} \lb \frac{\ell^2}{2E_o - 3}\,\frac{\p \psi^{\log}_{\mu\nu}}{\p E_o} \rb_{E_o = 3} =  \frac{c'\ell^4}{(2E_o - 3)^2}\,\frac{\p^2 \psi^{\log}_{\mu\nu}}{\p E_o^2}\bigg |_{E_o = 3} + \dots
  \end{align}
where we have dropped the term proportional to $\psi^{\log}_{\mu\nu}$. Using eqs.~\eqref{eq:einsteinmodes1} and \eqref{eq:einsteinmodes2} we find 
  \begin{align}
  h_{\mu\nu} = \frac{2\ell^4}{9} \lB it + \log\lb \sqrt{2}\,\cosh(r)\rb \rB^2 \psi_{\mu\nu},
  \end{align}
with $c' = 2$ chosen so that eq.~\eqref{eq:llaux1} is satisfied up to log modes.

Not surprisingly this solution is transverse and traceless as required by the gauge condition and the fact that there are no scalar modes at the tricritical point. Near the boundary $h_{\mu\nu} \sim r^2 e^{-r}$ or as expected $h_{\mu\nu} \sim z^2\log^2(z)$ in Poincar\'{e} coordinates. As before, we find modes with helicity $\pm 2$ obtained from eqs.~\eqref{eq:llaux1} and \eqref{eq:llaux2} with $k_{\mu\nu} = \psi_{\mu\nu}^E$, and spin-1 modes obtained from $k_{\mu\nu} = \bar{\n}_{(\mu}A_{\nu)}$. 

The solutions of tricritical gravity are thus 
  \begin{align}
  h_{\mu\nu} = \left\{	\begin{array}{lcl} 
	\psi_{\mu\nu}^E, && \textrm{massless spin 2} \\ 
	\psi_{\mu\nu}^{\log}, && \textrm{massless spin 2, massive spin 1} \\ 
	\psi_{\mu\nu}^{\log^2}, && \textrm{massless spin 2, massive spin 1} 
	\end{array} \right.
  \end{align}
where the modes $\bar{\n}_{(\mu}A_{\nu)}$ that do contribute to $k_{\mu\nu}$ and lead to spin-1 log and log$^2$ modes are pure gauge modes of $h_{\mu\nu}$ and may be gauged away\footnote{In other words, these are null states with zero energy as in Einstein gravity.}. As mentioned in the introduction critical gravities are conjectured to be dual to logarithmic CFTs on the boundary \cite{Grumiller:2008qz, Skenderis:2009nt, Grumiller:2009mw, Grumiller:2009sn, Alishahiha:2010bw}. This follows, for a critical gravity of any rank, by noticing that the auxiliary fields live in a reducible but indecomposable representation of the conformal algebra. To see this consider the generator of time translations in AdS$_4$ which corresponds to the generator of dilations in the boundary CFT$_3$. When acting on the modes of tricritical gravity it yields
  \begin{align}
  \L_{H_1} \psi_{\mu\nu} = 3\, \psi_{\mu\nu}, && \L_{H_1} \psi_{\mu\nu}^{\log} = 3\, \psi_{\mu\nu}^{\log} - \frac{2}{3} \psi_{\mu\nu} && \L_{H_1} \psi_{\mu\nu}^{\log^2} = 3\, \psi_{\mu\nu}^{\log^2} - \frac{2}{3} \psi_{\mu\nu}^{\log}.
  \end{align}
With $\psi_{\mu\nu}$ and $\psi_{\mu\nu}^{\log}$ normalized by -2/3 we obtain the Jordan form
  \begin{align}
  \L_{H_1} \( \begin{array}{c} 
           \psi_{\mu\nu} \\ 
           \psi_{\mu\nu}^{\log} \\ 
           \psi_{\mu\nu}^{\log^2} 
           \end{array} \)
         = \(\begin{array}{ccc} 
            3 & 0 & 0 \\
	    1 & 3 & 0 \\
  	    0 & 1 & 3
	    \end{array} \)
	    \( \begin{array}{c} 
            \psi_{\mu\nu} \\ 
            \psi_{\mu\nu}^{\log} \\ 
            \psi_{\mu\nu}^{\log^2} 
            \end{array} \),
  \end{align}
that characterizes a non-unitary logarithmic representation of rank 3 \cite{Gurarie:1993xq,Flohr:2001tj, Flohr:2001zs}. This behavior is mirrored by the dual operators on the boundary CFT where the operator dual to $\psi_{\mu\nu}$ has vanishing norm and the operators dual to $\psi_{\mu\nu}^{\log}$, $\psi_{\mu\nu}^{\log^2}$ are its logarithmic partners.


\section{A unitary truncation}

Tricritical gravity, like many other theories with higher derivatives, is perturbatively non-unitary. This is most easily seen by finding its inner product and showing that it is always possible to construct modes of negative norm unless we truncate the spectrum as described in \cite{Bergshoeff:2012sc}. This truncation yields a unitary theory only for critical gravities of odd rank like the tricritical gravity studied in this paper. In this section we construct the inner product of tricritical gravity and show that a unitary truncation is always possible at linear order in the equations of motion in the absence of matter fields.

In \cite{Porrati:2011ku} the inner product of theories with non-diagonal kinetic terms was constructed. There it was shown that for an action of the form 
  \begin{align}
  S = \frac{1}{2} \int dt \(\dot{q}^T L \dot{q} + \dots \),
  \end{align}
where $q$ is a vector and $L$ a symmetric matrix, the inner product between two positive-frequency modes $\vp$ and $\phi$ may be defined as follows
  \begin{align}
  \BraKet{\vp}{\phi} = i \vp^{*T}L\dot{\phi}.
  \end{align}
Their results can be applied to a higher-derivative theory when its action is written in terms of auxiliary fields. To second order the action of tricritical gravity (see eq.~\eqref{eq:auxiliaryaction}) is given by 
  \begin{align}
  \begin{split}
  S_3 = -\frac{\k}{2} \Int{d}{\bar{g}} & \lB \(\bar{\n}_\s f_{\mu\nu}\)^2 + \g_1 \( \bar{\n}_\s f \)^2 - \a_o\l f_{\mu\nu}^2 - \g_o\l f^2 \right. \\ 
  & + 4 \lb k^{\mu\nu}\bar{L}_{\mu\nu}^{\ph{\mu\nu}\s\rho}h_{\s\rho} - \( k^{\mu\nu}f_{\mu\nu} - kf \) \rb + \dots \Big \}, \label{eq:auxiliaryaction2}
  \end{split}
  \end{align}
where all contractions are with respect to the background metric and $(h, k, f)$ obey the linearized equations of motion \eqref{eq:laux1} - \eqref{eq:laux3}. In terms of the transverse and traceless modes the kinetic terms read
  \begin{align}
  S_3 = -\frac{\k}{2} \Int{d}{\bar{g}} & \bar{g}^{00} \,\(  \bar{\n}_0 f_{\mu\nu} \bar{\n}_0 f^{\mu\nu} + 2 \bar{\n}_0 k_{\mu\nu} \bar{\n}_0 h^{\mu\nu} \) + \dots
  \end{align}
where we have used eq.~\eqref{eq:Hlinear} and an integration by parts in the second term. Using eqs.~\eqref{eq:llaux1} - \eqref{eq:llaux3} the inner product is given by
  \begin{align}
  \begin{split}
  \BraKet{\vp}{\phi} = -\frac{i\k}{4} \Int{d-1}{\bar{g}}\, \bar{g}^{00}& \lb \vp^{*}_{\mu\nu} \bar{\n}_0 \( \bbox + 2 \ell^{-2}\)^2 \phi^{\mu\nu} + \( \bbox + 2 \ell^{-2}\)^2 \vp^{*}_{\mu\nu} \bar{\n}_0  \phi^{\mu\nu}\right. \\ 
  & \left. + \(\bbox + 2 \ell^{-2}\)\vp^{*}_{\mu\nu} \bar{\n}_0 \( \bbox + 2 \ell^{-2}\) \phi^{\mu\nu}\rb.
  \end{split}
  \end{align}
Thus the inner product among the modes of tricritical gravity $(\psi, \psi^{\log}, \psi^{\log^2})$ takes the form
  \begin{align}
  \BraKet{\vp}{\phi} = \( \begin{array}{ccc} 
			   0 & 0 & \e \\
			   0 & \e & \eta \\
			   \e & \eta^* & \z
   	 		   \end{array} \),  \label{eq:innerproduct}
  \end{align}
where $\eta$, $\z$ are non-zero complex numbers, and $\e$ is up to a positive constant equal to the inner product between two modes in Einstein's theory \eqref{eq:einsteinhilbert}. It is therefore positive between helicity-2 modes and zero otherwise. Hence the $\bar{
\n}_{(\mu}A_{\nu)}$ modes are null states as expected.

Notice that it is possible to construct states of the form $\Ket{\psi^E_{\mu\nu}} + c|\psi_{\mu\nu}^{\log^2}\rangle$ whose norm can be made negative by choosing $c$ appropriately. Hence the theory is not unitary even if the energies of all modes are positive. However, as pointed out in \cite{Bergshoeff:2012sc} by setting $k_{\mu\nu} = 0$ in the linearized equations of motion \eqref{eq:llaux1} - \eqref{eq:llaux3} it is possible to remove the $\log^2$ modes so that the inner product becomes
  \begin{align}
  \begin{split}
  \BraKet{\vp}{\phi} = -\frac{i\k}{4}\Int{d-1}{\bar{g}}\, \bar{g}^{00} & \lb \(\bbox + 2 \ell^{-2}\)\vp^{*}_{\mu\nu} \bar{\n}_0 \( \bbox + 2 \ell^{-2}\) \phi^{\mu\nu}\rb.
  \end{split}
  \end{align}
The modes of the truncated theory are now $(\psi^E, \psi^{\log})$ and their inner product is given by
  \begin{align}
  \BraKet{\vp}{\phi} = \( \begin{array}{cc} 
			   0 & 0 \\
			   0 & \e \\
   	 		   \end{array} \).  \label{eq:innerproduct2}
  \end{align}
We thus obtain a theory with the modes of (bi)critical gravity but with a positive-definite Hilbert space. Here the Einstein modes $\psi^E_{\mu\nu}$ are null states with zero energy so they can be modded out of the physical spectrum. In three dimensions tricritical gravity has only spin-1 $\log$ and $\log^2$ modes since the only solutions to \eqref{eq:llaux3} are pure gauge. Thus the inner product vanishes up to boundary terms and the truncation yields a trivial theory in the bulk where the vacuum is the only physical state. On the other hand, in four dimensions the only propagating degrees of freedom are the helicity $\pm 2$ $\log$ modes with the same norm as the gravitons in Einstein's theory. Seemingly, we have obtained a power-counting renormalizable, higher-derivative theory of gravity with a positive metric.


\section{Conserved charges and inconsistency of the truncation}

At second order in the perturbative expansion around the AdS background, the truncation of tricritical gravity $k_{\mu\nu} = 0$ is no longer a solution to the now inhomogeneous equations of motion. In principle it is possible to restrict the spectrum to the unitary subsector even in the interacting theory by means of a conservation law that prevents the $k_{\mu\nu}$ modes from being sourced. This is analogous to what happens in chiral gravity, whose solutions have vanishing left-handed Killing charges \cite{Li:2008dq, Maloney:2009ck}. In tricritical gravity we find a contradiction in the definition of the conserved charges that makes this not only impossible, but renders the truncation inconsistent at second order. That is, the theory suffers from a linearization instability and the only consistent truncation is that where both $k_{\mu\nu}$ and $f_{\mu\nu} $ vanish. This leaves us with a rather uninteresting theory where the only physical state is the vacuum.

Let us begin by expanding the fields to second order,
  \begin{align}
  g_{\mu\nu} \ra \bar{g}_{\mu\nu} + h_{\mu\nu}^{(1)} + h_{\mu\nu}^{(2)} && F_{\mu\nu} \ra f_{\mu\nu}^{(1)} + f_{\mu\nu}^{(2)} && K_{\mu\nu} \ra k_{\mu\nu}^{(1)} + k_{\mu\nu}^{(2)}, 
  \end{align}
where $(h^{(1)}, f^{(1)}, k^{(1)})$ satisfy the linearized equations of motion \eqref{eq:laux1} - \eqref{eq:laux3}. To second order eq.~\eqref{eq:aux3} tells us that
  \begin{align}
  \bar{L}_{\mu\nu}^{\ph{\mu\nu}\rho\s}k_{\rho\s}^{(2)} & = Q_{3\, \mu\nu} = \frac{1}{2} T^{(2)}_{\mu\nu} - \frac{1}{2}\lb \wtil{R}^{(1)}_{\mu\nu}\bar{g}^{\rho\s} - \wtil{R}^{(1)} \d^\rho_\mu\,\d^\s_\nu + 2 L_{\mu\nu}^{(1)\rho\s} \rb k_{\rho\s}^{(1)},  \label{eq:lllaux3}
  \end{align}
where $L_{\mu\nu}^{(1)\rho\s}$, $\wtil{R}^{(1)}_{\mu\nu}$, and $\wtil{R}^{(1)}$ are linear in $h_{\mu\nu}^{(1)}$.  The Bianchi identity guarantees that $Q_{3\,\mu\nu}$ is covariantly conserved with respect to the background. Hence the formalism of \cite{Abbott:1981ff} may be used to construct the conserved charges of any critical gravity. Using eq.~\eqref{eq:Hlinear} it is not difficult to show that eq.~\eqref{eq:lllaux3} may be written as
  \begin{align}
  Q_{3\,\mu\nu} = \bar{\n}^\rho \bar{\n}^\s \K_{\mu\rho\nu\s} + \frac{\l}{d-1}\,\bar{g}^{\rho\s} \K_{\mu\rho\nu\s},  \label{eq:ADM}
  \end{align}
where 
  \begin{align}
  \K_{\mu\rho\nu\s}  & = \frac{1}{2} \( \bar{g}_{\mu\s} \H_{\nu\rho} + \bar{g}_{\nu\rho} \H_{\mu\s} - \bar{g}_{\mu\nu} \H_{\rho\s} - \bar{g}_{\rho\s} \H_{\mu\nu} \), \\
  \H_{\mu\nu}  & = k^{(2)}_{\mu\nu} - \frac{1}{2}\,\bar{g}_{\mu\nu}k^{(2)}.
  \end{align}
Denoting by $\xi$ a Killing vector and by $\eta_{\mu} = (1,0,0,0)$ the \emph{vector} normal to a spacelike hypersurface, the Killing charges are given by
  \begin{align}
  Q[\xi] = \Int{d-1}{\bar{g}} \,Q_{3\,\mu\nu} \xi^\nu \eta^\mu  & = \Int{d-1}{\bar{g}} \, \n_\rho \( \bar{\n}_{\s} \K^{\mu\rho\nu\s} \xi_\nu- \K^{\mu\s\nu\rho} \bar{\n}_\s \xi_\nu \) \eta_\mu,
  \end{align}
where we have used the fact that $\K_{\mu\rho\nu\s}$ has the symmetries of the Riemann tensor and $\bar{\n}_\rho \bar{\n}_\s \xi_{\nu} = \bar{R}^\l_{\phantom{\l}\rho\s\nu} \xi_\l$. The conserved surface charges are thus
  \begin{align}
  Q[\xi] = \oint dS_i \sqrt{|\bar{g}|}  \, \( \bar{\n}_{\s} \K^{0 i \nu\s} \xi_\nu- \K^{0 j \nu i} \bar{\n}_j \xi_\nu \).
  \end{align}

In the truncated theory $k^{(1)}_{\mu\nu} = k^{(2)}_{\mu\nu} = 0$ so the conserved charges vanish. With $f^{(1)}_{\mu\nu} \ne 0$, this leads to a contradiction as evidenced by the fact that the bulk expression for the energy is of definite sign. To see this consider the on-shell expression for $Q_{3\,\mu\nu}$ which in the truncated theory is given by one-half the on-shell effective stress-energy tensor. The latter is obtained from variation of eq.~\eqref{eq:auxiliaryaction} with respect to the metric where we ignore the $\d H_{\rho\s}/\d g^{\mu\nu}$ term; it may be conveniently written as
  %
  %
  %
  \begin{align}
  T^{(2)}_{\mu\nu}  = -\frac{\bar{g}_{\mu\nu}}{8} \bbox \lb f^{(1)}_{\rho\s}f^{(1)\rho\s} \rb + \frac{1}{2} \bar{\n}_\mu f^{(1)}_{\rho\s} \bar{\n}_\nu f^{(1)\rho\s} - \frac{1}{2}\bar{\n}_\rho \lb  f^{(1)}_{\mu\s}\bar{\n}_\nu f^{(1)\rho\s} - f^{(1)\rho\s}\bar{\n}_\mu f^{(1)}_{\nu\s}  + (\mu \leftrightarrow \nu)\rb, \label{eq:tmunuonshell2}
  \end{align}
where we have used eqs.~\eqref{eq:traceless} and \eqref{eq:llaux2} with $k_{\mu\nu} = 0$. The volume integral expression for the energy is thus given by
  \begin{align}
  Q_E = \frac{1}{2} \Int{d-1}{\bar{g}} \,T^{(2)}_{\mu\nu} \xi_E^\mu \eta^\nu, && \xi_E^\mu = (1,0,0,0).
  \end{align}
Since the energy is conserved in time we can turn it into a spacetime integral to obtain
  \begin{align}
  \begin{split}
  Q_E = \lim_{T \ra \infty} \frac{1}{8T} \Int{d}{\bar{g}} & \lB  -\frac{1}{4} (f^{(1)}_{\rho\s})^2\, \bbox (\xi_E^\mu \eta_\mu) + \bar{\n}_\mu f^{(1)}_{\rho\s} \bar{\n}_\nu f^{(1)\rho\s} \xi_E^\mu \eta^\nu \right. \\
   & \phantom{\bigg{\{}} +  \lb  f^{(1)}_{\mu\s}\bar{\n}_\nu f^{(1)\rho\s} - f^{(1)\rho\s}\bar{\n}_\mu f^{(1)}_{\nu\s}  + (\mu \leftrightarrow \nu) \rb \bar{\n}_\rho (\xi_E^\mu \eta^\nu)  \bigg \}.
  \end{split}
   \end{align}
In global coordinates (see eq.~\eqref{eq:globalmetric}) in $d \ge 3$ the only non-vanishing Christoffel symbols with a 0 index are
  \begin{align}
  \bar{\gg}^r_{00} = - \frac{1}{2} \, \bar{g}^{rr}\p_r \bar{g}_{00}, && \bar{\gg}^0_{r 0} = \bar{\gg}^0_{0 r} = \frac{1}{2} \, \bar{g}^{00} \p_r \bar{g}_{00}, && \bar{\gg}^0_{r 0} = - \frac{\bar{g}^{00}}{\bar{g}^{rr}}\bar{\gg}^r_{00},
  \end{align}
where $r$ is the radial coordinate. Using the expressions for $\xi_E$ and $\eta$, and 
  \begin{align}
  \bbox (\xi_E^\mu \eta_\mu) = 0, && \bar{\n}_\rho (\xi_E^\mu\eta^\nu) = \bar{g}^{\nu 0} \bar{\gg}^\mu_{\rho 0} - \bar{g}^{\nu\g} \bar{\gg}^{0}_{\rho \g} \d_0^\mu,
  \end{align}
we thus obtain
  \begin{align}
  Q_E & = \lim_{T \ra \infty} \frac{1}{8T} \Int{d}{\bar{g}} \bar{g}^{00}\lb \bar{\n}_0 f^{(1)}_{\mu\nu} \bar{\n}_0 f^{(1)\mu\nu} - 2  \bar{g}^{rr} \bar{\gg}^{0}_{r 0} \( f^{(1)}_{r \s} \bar{\n}_0 f_0^{(1)\s} - f_0^{(1) \s} \bar{\n}_0 f^{(1)}_{r \s} \) \rb, \\
  Q_E & = \lim_{T \ra \infty} \frac{1}{8T} \Int{d}{\bar{g}} \,\bar{g}^{00} \p_0 f^{(1)}_{\mu\nu} \bar{\n}_0 f^{(1)\mu\nu}. 
  \end{align}
Now let $f^{(1)}_{\mu\nu} = \vp_{\mu\nu} + \vp_{\mu\nu}^*$ where $\vp_{\mu\nu}$ is a positive-frequency mode, i.e. $\vp \sim e^{-i\omega t}$ where $\omega > 0$. Then the terms proportional to $\vp_{\mu\nu}\vp^{\mu\nu}$ and $\vp_{\mu\nu}^* \vp^{*\mu\nu}$ drop out and the energy is given by
  \begin{align}
  Q_E = \frac{i\omega}{2} \Int{d-1}{\bar{g}} \bar{g}^{00} \vp^*_{\mu\nu} \bar{\n}_0 \vp^{\mu\nu},
  \end{align}
which is proportional to the norm in Einstein's theory and of definite negative sign. This comparison is appropriate since $f^{(1)}_{\mu\nu}$ in the truncated theory is the solution to the gauge-fixed linearized equations of motion of Einstein's theory in an AdS background. That is, it obeys 
  \begin{align}
  (\bbox + 2\ell^{-2}) f^{(1)}_{\mu\nu} = 0, && \bar{\n}^\mu f^{(1)}_{\mu\nu} = 0, && f^{(1)} = 0.
  \end{align}
Thus the bulk expression for the Killing energy is non-vanishing in the unitary subsector, and cannot be made to vanish by an arbitrary superposition of modes, in contrast to the vanishing of the surface charge. {\em This is a contradiction so the truncation $k_{\mu\nu}$ = 0 is inconsistent at second order.}


\section{Generic parity-even tricritical gravities}

The inconsistency shown in the previous section is not unique to the minimal tricritical gravity described by eqs.~\eqref{eq:tricriticalaction2}, \eqref{eq:auxiliaryaction}. To see this recall that we can always add to the action terms cubic and higher order in curvature tensors as long as these do not contribute to the linearized equations of motion. We now show that their contribution to the bulk energy vanishes so parity-even tricritical gravities in three and four dimensions cannot be made unitary by the truncation of \cite{Bergshoeff:2012sc}. Let us begin by considering the possible terms we may add to the action \eqref{eq:auxiliaryaction} that contribute to the second-order, effective stress-energy tensor. This immediately rules out terms cubic in $F_{\mu\nu}$ which contribute at third order. Also, since we are interested in the on-shell expression for $T^{(2)}_{\mu\nu}$ in the truncated theory we may omit terms that contain $K_{\mu\nu}$. Thus the terms that we may add to the action are of the form
  \begin{align}
  S'_{3} = -\k \Int{d}{\bar{g}} \( - a_1 \wtil{R} \O - a_2 \wtil{R}_{\mu\nu}\O^{\mu\nu} + a_3 \wtil{R}_{\mu\rho\nu\s} \O^{\mu\rho\nu\s}  \),  \label{eq:tricriticalaction4}
  \end{align}
where $a_1, a_2, a_3$ are free parameters and $\O, \O_{\mu\nu}, \O_{\mu\rho\nu\s}$ are quadratic in $F_{\mu\nu}$ and $\wtil{R}, \wtil{R}_{\mu\nu}, \wtil{R}_{\mu\rho\nu\s}$ with the curvature terms vanishing in the background, e.g. $\wtil{R} = R - \bar{R}$. For example, the $\O$ terms may be given by
  \begin{align}
  \O = F_{\mu\nu}F^{\mu\nu}, && \O_{\mu\nu} = F_{\mu\s}F^{\s}_{\ph{\s}\nu},  && \O_{\mu\rho\nu\s} = F_{\mu\nu}F_{\rho\s},
  \end{align}
among many other possibilities. An exhaustive list of cubic curvature terms is given in ref.~\cite{Myers:2010ru}.

It is always possible and convenient to make $\O_{\mu\nu}$ and $\O_{\mu\rho\nu\s}$ obey the same algebraic symmetries as the curvature tensors with which they are contracted. Then the contribution of eq.~\eqref{eq:tricriticalaction4} to the second-order, on-shell, effective stress-energy tensor is given by\footnote{Clearly if the $\O$ terms depend on curvature tensors this expression contains extra terms with the same structure, i.e. $\O \ra \O'$ and $a_i \ra a'_i$, so the results are unchanged.}
  \begin{align}
  \begin{split}
  T'^{(2)}_{\mu\nu} = & a_1 \( \bar{\n}_\mu \bar{\n}_\nu - \bar{g}_{\mu\nu} \bbox - \bar{g}_{\mu\nu} \l \) \O^{(2)} + 2a_3 \( \bar{\n}^\rho \bar{\n}^\s + \frac{\l}{d-1} \bar{g}^{\rho\s} \) \O^{(2)}_{\mu\rho\nu\s} \\
  & + \frac{a_2}{2} \( \bar{\n}^\rho \bar{\n}_\mu \,\d_\nu^\s + \bar{\n}^\rho \bar{\n}_\nu \,\d_\mu^\s - \bbox \,\d_\mu^\rho \,\d_\nu^\s - \bar{g}_{\mu\nu}\bar{\n}^\rho \bar{\n}^\s - 2 \l \,\d_\mu^\rho \,\d_\nu^\s \) \O^{(2)}_{\rho\s} \label{eq:tmunu2}
  \end{split}
  \end{align}
where it is understood that $\O^{(2)}, \O^{(2)}_{\mu\nu}$, and $\O^{(2)}_{\mu\rho\nu\s}$ are quadratic in the first-order perturbations $h^{(1)}_{\mu\nu}$ and $f^{(1)}_{\mu\nu}$. It is not difficult to show that $T'^{(2)}_{\mu\nu}$ is given by eq.~\eqref{eq:ADM} where
  \begin{align}
  \K_{\mu\rho\nu\s} \ra \K_{\mu\rho\nu\s} + 2a_3 \O^{(2)}_{\mu\rho\nu\s}, && \H_{\mu\nu} = a_2 \O^{(2)}_{\mu\nu} + a_1 \bar{g}_{\mu\nu}\O^{(2)},
  \end{align}
so that its contribution to the bulk energy vanishes
  \begin{align}
  Q_E = \frac{1}{2}\Int{d-1}{\bar{g}} \,T'^{(2)}_{\mu\nu}\xi_E^{\mu}\eta^{\nu} = \frac{1}{2}\Int{d-1}{\bar{g}} \, \n_\rho \( \bar{\n}_{\s} \K^{\mu\rho\nu\s} \xi_\nu- \K^{\mu\s\nu\rho} \bar{\n}_\s \xi_\nu \) \eta_\mu = 0.
  \end{align}

This means that we cannot make the truncation $k_{\mu\nu} = 0$ consistent by the addition of cubic or higher order terms to the action. Thus parity-even tricritical gravity in three and four dimensions has no non-trivial positive-metric subspace.


\section{Conclusions}
In this paper we have constructed the simplest 6-derivative action in $d$ dimensions that realizes the degenerate scalar toy model of ref.~\cite{Bergshoeff:2012sc} for a theory of gravity. Although the theory contains a subspace with a positive-definite inner product in the linearized approximation we see that this truncation is no longer consistent at second order, a possibility already mentioned in ref~\cite{Bergshoeff:2012ev}. This inconsistency is present in any parity-even tricritical gravity in three and four dimensions described by the minimal action given in eq.~\eqref{eq:tricriticalaction2} and any number of additional cubic or higher-order curvature terms that do not affect the linearized equations of motion. Our results are consistent with the analysis of \cite{Bergshoeff:2012ev}, where tricritical gravity was studied in three dimensions. There it was found that when the $\log^2$ modes vanish, the conserved charges of the boundary CFT also vanish. This is what we also find in $d$ dimensions, but we also find that the condition of vanishing charge does not consistently select a unitary subspace.

Nevertheless it is still possible that other tricritical gravities may  allow for a consistent truncation to a positive-definite Hilbert space. In three dimensions there exists a class of {\em parity-odd} tricritical gravities, where the terms $(\n_\s H_{\mu\nu})^2$ and $(\n_\s H)^2$ in \eqref{eq:tricriticalaction} are replaced by the Lorentz-Chern-Simons term and the coefficients $\a_n, \b_n$ also change \cite{Bergshoeff:2012ev}. In dimensions greater than four another class of tricritical gravities exists which contains the Riemann tensor squared, an example of which has been recently considered in \cite{Nutma:2012ss}. It would be interesting to check whether the bulk expression for the conserved charges is of definite sign in these theories.

In ref.~\cite{Bergshoeff:2012sc} it was shown that a unitary truncation at linear order exists for any higher-derivative theory of odd rank. The minimal tricritical gravity studied here is the simplest example where the truncation works, albeit only to linear order. Although it seems unlikely, it is possible that the truncation is consistent at second and higher orders in theories of higher rank. Even if this turns out to be the case, we cannot restrict multicritical gravities to the positive-metric subspace by means of the conserved charges given by Gauss's law. In these theories the second order equations of motion are the analog of eqs.~\eqref{eq:charges}, namely
  \begin{align}
  L_{\mu\nu}^{\phantom{\mu\nu}\rho\s}h^{(2)}_{\rho\s} & = f_{\mu\nu}^{(2)} + Q_{1\,\mu\nu}[h^{(1)},f^{(1)},\dots,k^{(1)}], \notag \\
  \phantom{L_{\mu\nu}^{\phantom{\mu\nu}\rho\s}f^{(2)}_{\rho\s}} & \:\:\vdots \phantom{k_{\mu\nu}^{(2)} + Q_2[h^{(1)},f^{(1)},k^{(1)}]} \notag \\
  L_{\mu\nu}^{\phantom{\mu\nu}\rho\s}k^{(2)}_{\rho\s} & = Q_{3\, \mu\nu}[h^{(1)},f^{(1)},\dots,k^{(1)}]. \label{eq:charges2}
  \end{align}
By setting  $k_{\mu\nu} = 0$ in eq.~\eqref{eq:charges2}, one makes the corresponding charge  $Q_E=\Int{d-1}{\bar{g}}\bar{g}^{00} Q_{3\,\mu 0}\xi^\mu$ vanish. This condition can restrict to a unitary subspace only if it sets $(r-1)/2$ modes to zero (namely, $\psi^{\log^{r-1}}, ..., \psi^{\log^{ (r+1)/2}}$,
see ref.~\cite{Bergshoeff:2012sc}). This cannot be achieved by setting $Q_E=0$ because the charge density  of the modes $\psi^{\log^{r-1}}, ..., \psi^{\log^{ (r+1)/2}}$ is not positive definite. 

We cannot exclude at the moment that other, parity-odd tricritical or multicritical gravities admit a consistent truncation. Moreover, the potential applications of parity-even and parity-odd multicritical gravities to condensed matter physics via the AdS/CFT duality remain to be explored. This makes further study of these theories worthwhile.



\subsection*{Acknowledgments}
We would like to thank  Eric A. Bergshoeff, Sjoerd de Haan, Wout Merbis, Jan Rosseel and Thomas Zojer for making preprint arXiv:1206.3089 [hep-th] available to us prior to posting. M.P. is supported in part by NSF grant PHY-0758032, and by ERC Advanced Investigator Grant n.226455 {\em Supersymmetry, Quantum Gravity and Gauge Fields (Superfields)}. M.P. would like to thank ICTP for its hospitality during the completion of this paper.



\end{document}